\documentclass[conference]{IEEEtran}
\IEEEoverridecommandlockouts
\usepackage{cite}
\usepackage{amsmath,amssymb,amsfonts}
\usepackage{algorithm}
\usepackage{graphicx}
\usepackage{textcomp}
\usepackage{xcolor}
\usepackage{algpseudocode}

\graphicspath{{figures/}{./}}

\newtheorem{remark}{Remark}
\newtheorem{assumption}{Assumption}
\newcommand{\T}{\mathbb{T}}
\newcommand{\p}{\mathbb{P}}
\newcommand{\Fl}{\mathrm{Fl}}
\newcommand{\ind}{\mathbf{1}}
\usepackage{tikz}
\usepackage{hyperref}

\newcommand\copyrightnotice{%
\begin{tikzpicture}[remember picture,overlay]
\node[anchor=south,yshift=20pt] at (current page.south)
  {\parbox{0.8\textwidth}{\centering\footnotesize
  This work has been submitted to the IEEE for possible publication. Copyright may be transferred without notice, after which this version may no longer be accessible.}};
\end{tikzpicture}}

\def\BibTeX{{\rm B\kern-.05em{\sc i\kern-.025em b}\kern-.08em
    T\kern-.1667em\lower.7ex\hbox{E}\kern-.125emX}}
\begin{document}

\title{LEO Constellation Analysis for Satellite IoT: An Analysis of Coverage Gaps and Retransmissions\\
\thanks{This research was funded in part by the European Union’s Horizon Europe research and innovation programme under the Marie Skłodowska-Curie grant agreement No. 101282240.}
}

\author{
\IEEEauthorblockN{Lowie Rombouts, Jeroen Famaey and Roger Sanchez-Vital}

\IEEEauthorblockA{\textit{IDLab, University of Antwerp -- imec}, Antwerp, Belgium\\
\{lowie.rombouts, jeroen.famaey, roger.sanchezvital\}@uantwerpen.be}
}

\maketitle

\copyrightnotice

\begin{abstract}
Low Earth Orbit (LEO) constellations are a promising way to connect Internet of Things (IoT) devices beyond terrestrial coverage, yet the intermittent visibility of their satellites imposes hard constraints on system and protocol design. Coverage probability and other snapshot statistics, which are often computed without accounting for the terminal's latitude, are insufficient for this purpose. Designers also need the temporal structure of coverage: the rate of coverage gaps, their duration distribution and mean, and quantities such as the probability that a retransmission succeeds after an initial transmission has failed. We develop an ergodic framework for Walker Star constellations in which these quantities reduce to spatial integrals over a two-dimensional torus that can be evaluated efficiently. We validate the framework against a seven-day Simplified General Perturbations 4 (SGP4) simulation of the Iridium NEXT constellation and find close agreement for the coverage probability, the revisit-time distribution, and the retransmission success probability. A comparison of Walker Star configurations further shows that constellations with similar satellite counts can yield substantially different packet-delivery probabilities with at most one retransmission, depending on the number of orbital planes. The results show that the ergodic model accurately gives the long-run values that finite time-domain simulations can only approximate, providing researchers with a tractable basis for the analytical modelling of IoT connectivity over non-terrestrial networks (NTN).
\end{abstract}

\begin{IEEEkeywords}
LEO satellite constellations, satellite IoT, ergodic theory, coverage analysis, revisit time,
retransmission success
\end{IEEEkeywords}

\section{Introduction}
In 2025, more than 21 billion Internet of Things (IoT) devices were active worldwide, and this number is expected to nearly double by 2030 \cite{iotanalytics2025}. Sensors for smart agriculture, wildlife tracking, and remote weather stations are being deployed at scale. Many of them operate beyond the reach of any terrestrial network, and for these devices an alternative is essential. Low Earth Orbit (LEO) constellations have become a promising solution. LEO satellites usually orbit at altitudes between 400~km and 1200~km, which yields path losses and latencies that can be compatible with the simple radios of IoT devices. A rapidly growing body of standardization work has followed, and 3GPP has supported NB-IoT over NTN \cite{figaro2026ntn} since Release~17. Although several studies, such as \cite{kodheli2018, testi2025, fraire2026blind}, are valuable from an engineering perspective, they lack a mathematical description of the constellation that captures all the essential temporal statistics, which limits how well protocols can be evaluated.

The main practical challenge is that LEO coverage is intermittent by nature for the sparse, cost-effective constellations commonly used for IoT. A ground terminal therefore does not see satellites continuously. Instead, it sees a sequence of visibility windows separated by gaps that, depending on the constellation and latitude, can last from minutes up to hours. For low-power devices that cannot afford to listen continuously to find out whether a satellite is overhead, these gaps strongly constrain system performance. System design therefore has to answer several temporal questions: What is the distribution of the gaps? How long, on average, must a device wait after a failed transmission? If a packet is retransmitted after a fixed delay, how does the probability of success vary with time, and which retransmission delay is optimal? The coverage probability at a random instant cannot answer these questions.

Stochastic geometry has been a dominant analytical tool for satellite communication modelling in recent years. By treating satellites as random point processes, often binomial or Poisson, one can obtain elegant (semi-)analytical expressions for the coverage probability and the carrier-to-interference-plus-noise ratio (CINR) \cite{talgat2021, okati2020, dong2025}. This approach has produced numerous results and has been extended to IoT uplink scenarios \cite{talgat2024}. Its limitation is structural: real constellations such as Iridium, Starlink or OneWeb are not random. They are carefully designed Walker Star or Walker Delta configurations, and the Poisson approximation misses this geometry completely. Cox point processes, recently applied to satellite networks \cite{choi2025novel}, are a partial solution: they place the satellites on orbits, but the positions on these orbits remain random. These works are almost always validated against a Monte Carlo simulation of the same random model, not against propagated orbits of real constellations.

More constellation-aware studies are available. Ulybyshev \cite{Ulybyshev2015} developed a geometric method to compute the maximum revisit time, i.e., the worst-case gap duration. More recently, two works showed how the ergodicity of a linear flow on a torus leads to an analytical expression for the coverage probability \cite{Graven2022Rapid, choi2025walker}. Both works, however, stop at instantaneous statistics. Neither addresses the temporal structure of coverage gaps, which is exactly what satellite IoT system and protocol design needs. This paper closes that gap and shows that the ergodic framework extends naturally to the statistics that govern satellite IoT performance, such as the gap rate, the gap-duration distribution, and the retransmission success probability, each expressed as an integral over the same torus.

We validate the coverage probability, revisit-time CDF, and retransmission success probability against an independent seven-day Simplified General Perturbations 4 (SGP4) simulation of Iridium NEXT. We then compare Walker Star configurations with different numbers of planes and satellites to illustrate how the framework informs constellation design.

The remainder of the paper is organized as follows. Section~\ref{sec:model} introduces the Walker Star system model and its torus state space. Section~\ref{sec:ergodic} develops the ergodic framework and derives the coverage statistics. Section~\ref{sec:numerical} presents the numerical results and the SGP4 validation for Iridium NEXT, and Section~\ref{sec:conclusion} concludes the paper.

\section{System Model} \label{sec:model}
While the methods in this paper generalize to many types of constellations, we base our system model on a Walker Star, a common design choice for IoT coverage. Walker Star constellations are described by the Walker notation $i\!:\!T/P/F$, where $i$ is the inclination, $T$ the total number of satellites, $P$ the number of orbital planes, and $F$ the phasing parameter. We adopt the following assumption.
\begin{assumption}\label{ass:twobody}
    The Earth is modelled as a rigid sphere of radius $r_E=6371$~km with a spherically symmetric mass distribution, rotating at a constant angular velocity $\omega_E$ about its polar axis. The satellites stay at a constant altitude $h$. Under this assumption, the orbital motion reduces to the classical two-body (Keplerian) problem.
\end{assumption} All satellites move on circular orbits of common radius $r_S=h+r_E$, and each plane contains $S=T/P$ satellites. Unlike in a Walker Delta constellation, where the planes are spread evenly over $360^\circ$, the planes of a Walker Star are spread evenly over $180^\circ$. In an Earth-fixed frame,
the longitude of the ascending node $\Omega_j$ of plane $j$ and the argument of latitude $u_{jk}$ of satellite $(j,k)$ are
\begin{equation}\label{eq:motionsat}
    \Omega_j(t)=\omega(t)+\frac{j\pi}{P},\qquad
    u_{jk}(t)=u(t)+\frac{2\pi k}{S}+j\Delta,
\end{equation}
where $j \in \{0,\dots,P-1\}$, $k \in \{0,\dots,S-1\}$, and $\Delta$ is the inter-plane phase offset. Common choices are $\Delta=2\pi F/T$, as in Walker Delta constellations, and $\Delta=\pi/S$.

\begin{figure}[t]
  \centering
  \includegraphics[width=\columnwidth]{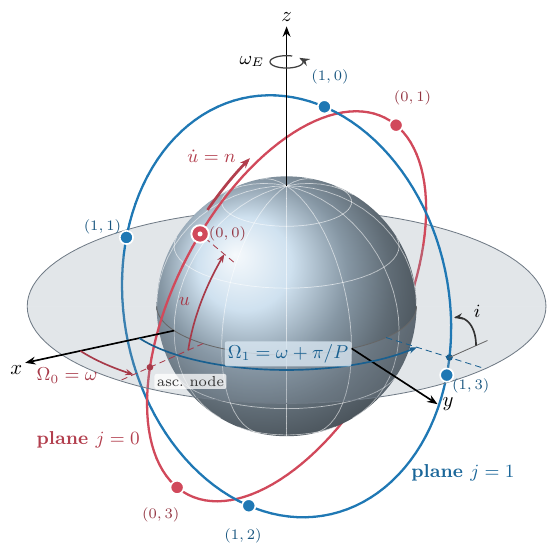}
  \caption{Geometry of the Walker Star constellation $80^\circ\!:\!8/2/1$
    ($P=2$, $S=4$, $\Delta=\pi/S$) in the Earth-fixed frame; $\Omega_j$ is measured from the Greenwich meridian ($x$-axis) and $u_{jk}$ from the ascending node.}
  \label{fig:walker_star}
\end{figure}

The position of satellite $(j,k)$ in the Earth-fixed frame is
\begin{equation}\label{eq:position}
  X_{jk} = r_S
  \begin{pmatrix}
    \cos\Omega_j\cos u_{jk} - \sin\Omega_j\sin u_{jk}\cos i\\[2pt]
    \sin\Omega_j\cos u_{jk} + \cos\Omega_j\sin u_{jk}\cos i\\[2pt]
    \sin u_{jk}\sin i
  \end{pmatrix}.
\end{equation}

Under Assumption~\ref{ass:twobody}, the motion of satellite $(0,0)$ is
\begin{equation}\label{flow}
    \omega(t)=\omega_0-\omega_E t,\qquad
    u(t)=u_0+nt,
\end{equation}
where $n$ is the mean motion and $(\omega_0,u_0)$ is the initial state; by
\eqref{eq:motionsat}, it determines the motion of all satellites. The positions
\eqref{eq:position} are invariant under $\omega\to\omega+2\pi$. Moreover, the
shift $u\to u+2\pi/S$ maps satellite $(j,k)$ to satellite $(j,k+1 \bmod S)$ and
therefore leaves the set of satellite positions unchanged. Since an observer is
concerned only with the number of visible satellites, not with which satellites
are visible, the state space of satellite $(0,0)$ reduces to the torus
$\T=\mathbb R^2/\Lambda_*$, where the lattice of the Walker Star is
$\Lambda_*=2\pi\mathbb Z\times (2\pi/S)\mathbb Z$.

\begin{figure}[!t]
  \centering
  \includegraphics[width=0.95\columnwidth]{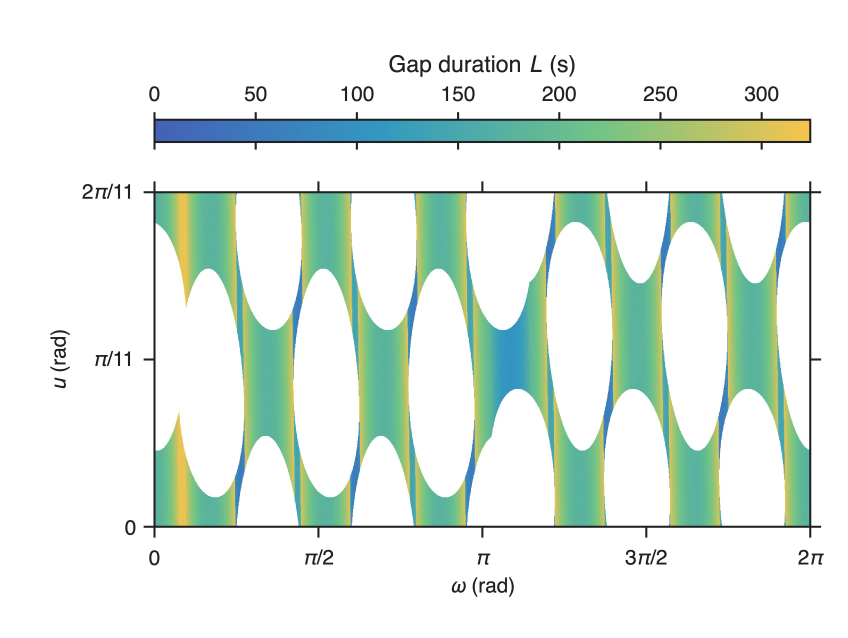}
  \caption{Coverage set $C$ (white) and gap set $G$ (coloured) on the torus
  $\T$, shown over one fundamental domain
  $[0,2\pi)\times[0,2\pi/S)$, for the Iridium constellation and the 
  terminal in Antwerp (parameters in Table~\ref{tab:params}). The colour of
  each state $x\in G$ indicates the duration $L(x)$ of the gap
  containing it, in seconds.}
  \label{fig:coverage_gap_antwerp}
\end{figure}

Fix an observer at $p\in\mathbb R^3$ with $\lVert p\rVert = r_E$. The elevation $\varepsilon$ of a satellite at position $X$
seen from $p$ satisfies
\begin{equation}
  \sin\varepsilon(X) = \frac{\langle X-p,\  p\rangle}{\lVert p\rVert\lVert X-p\rVert}.
\end{equation}
For a minimum elevation (mask) angle $\varepsilon_{\min}$, a state
$x=(\omega,u)\in\T$ provides coverage if at least one satellite is above the
mask. We define the \emph{coverage set} $C$ and the \emph{gap set} $G$ as
\begin{equation}\label{eq:CG}
  C = \Bigl\{x\in\T : \max_{j,k}\ \varepsilon\bigl(X_{jk}(x)\bigr) \ge \varepsilon_{\min}\Bigr\},
  \qquad
  G = \T\setminus C .
\end{equation}
The map $x\mapsto\varepsilon(X_{jk}(x))$ is continuous;
hence $C$ is closed and $G$ is open. Fig.~\ref{fig:coverage_gap_antwerp}
illustrates both sets for the Iridium configuration of Section~\ref{sec:numerical}
and an observer in Antwerp, Belgium.

\begin{remark}
    All quantities in this article depend on the observer location $p$, which we
    do not denote explicitly to ease notation. If the ground track does not
    repeat ($\omega_E/n\notin\mathbb Q$, see Section~\ref{sec:ergodic}), they
    depend only on the latitude of $p$, since a change of the observer's
    longitude corresponds to a translation in $\omega$. Almost all constellations satisfy this condition.
\end{remark}

\section{Ergodic Analysis}\label{sec:ergodic}
This section develops an ergodic framework for the coverage statistics that
determine how well a constellation meets a set of requirements. The state of
the constellation evolves as a linear flow on the torus $\T$, and when
this flow is ergodic, long-term time averages along a single trajectory equal
spatial averages over $\T$. Applied to the indicator of the
coverage set, this approach yields the coverage probability, as shown
in~\cite{Graven2022Rapid} and~\cite{choi2025walker}. The coverage probability,
however, describes the constellation only at a single random instant and says
nothing about how coverage evolves over time: two constellations with the same
coverage probability may have very different gap durations. These temporal
properties govern the performance of IoT protocols. The quantities treated in this section are not exhaustive; other statistics, such as the contact duration, can be treated in the same way. The main goal of this section is to show how ergodic theory makes it possible to model temporal statistics in semi-analytical form.

\subsection{Preliminaries from Ergodic Theory}
We use the following concepts from ergodic theory~\cite{cornfeld1982ergodic}.
The measure space $(\T, \mathcal{B}(\T), \mu)$ consists of the torus $\T$
equipped with its Borel $\sigma$-algebra and its normalized Lebesgue measure
$d\mu=\frac{S}{4\pi^2}\,d\omega\, du$. The equations of motion \eqref{flow}
define the flow $\Fl_t(x)=x+t(-\omega_E,n) \bmod \Lambda_*$, for
$(t,x)\in\mathbb R\times\T$, in the sense
of~\cite{cornfeld1982ergodic}. Since $\Fl_t$ acts by
translations, it preserves $\mu$. By the Birkhoff--Khinchin
theorem~\cite{cornfeld1982ergodic}, for every
$f\in L^1(\T,\mu)$ the time average
\begin{equation}\label{eq:timeaverage}
    \lim_{H\to \infty}\frac{1}{H}\int_0^H f\bigl(\Fl_t(x)\bigr)\,dt
\end{equation}
exists for $\mu$-almost all $x\in \T$. If $\omega_E/n\in \mathbb R\setminus \mathbb Q$,
the measure space together with the flow is an ergodic dynamical
system~\cite{cornfeld1982ergodic}, and the time average
equals the spatial average:
\begin{equation}\label{eq:spatialaverage}
    \lim_{H\to \infty}\frac{1}{H}\int_0^H f\bigl(\Fl_t(x)\bigr)\,dt=\int_\T f\,d\mu .
\end{equation}
%This is a well-known property of linear flows on tori; see the chapter on translations on the torus~\cite[Ch.~3, \S1]{cornfeld1982ergodic}.

Applied to $f=\ind_C$, \eqref{eq:spatialaverage} yields the long-term fraction
of time during which an observer has coverage, from which coverage
probabilities can be derived. For a device that transmits at a uniformly random time over a finite horizon $H$,
$V\sim \mathrm{U}[0,H]$, the coverage probability is
\begin{equation}\label{eq:covprob}
    \p\bigl(\Fl_V(x_0)\in C\bigr)=\frac{1}{H}\int_0^H\ind_C\bigl(\Fl_t(x_0)\bigr)\,dt
    \xrightarrow{H\to \infty}\mu(C)
\end{equation}
for $\mu$-almost all initial states $x_0\in \T$. In this work, coverage is the only requirement for a successful transmission. Under this assumption, the coverage probability coincides with the probability of successful transmission for a randomly transmitting device. All limits in the remainder of
this section hold for $\mu$-almost all $x_0$, and for every $x_0\in\T$ when the
integrand is Riemann integrable.

\subsection{Ergodic Estimation of Coverage Statistics}
For $x\in G$, we define the time until coverage is restored, $\tau^+(x)$, and the time elapsed since coverage was lost, $\tau^-(x)$, by
\begin{equation}
\begin{aligned}
    \tau^+(x)&= \inf\{t\geq0:\Fl_t(x)\in C\},\\
    \tau^-(x)&= \inf\{t\geq0:\Fl_{-t}(x)\in C\}.
\end{aligned}
\end{equation}
\textbf{Gap Distribution.}
The duration of the gap containing $x\in G$, i.e., the revisit time, is then
\begin{equation}
    L(x)=\tau^+(x)+\tau^-(x),
\end{equation}
and the maximum revisit time is
\begin{equation}\label{eq:maxrevisit}
    \sup_{x\in G}L(x).
\end{equation}
Solving \eqref{eq:maxrevisit} is similar in spirit to the approach
in~\cite{Ulybyshev2015}. Whereas the maximum revisit time captures only the
worst case, applications such as satellite IoT also depend on the mean and the
distribution of the revisit time, which the ergodic framework provides.

Let $N(H)$ denote the number of coverage gaps along the trajectory
$(\Fl_t(x_0))_{0\le t\le H}$. Since a gap of duration $\ell=b-a$ contributes
$\int_a^b (1/\ell)\,dt = 1$ to the integral
\begin{equation}\label{eq:countgap}
    \tilde N(H)=\int_0^H \frac{\ind_G\bigl(\Fl_t(x_0)\bigr)}{L\bigl(\Fl_t(x_0)\bigr)}\,dt,
\end{equation}
$\tilde N(H)=N(H)$ whenever there is coverage at $t=0$ and $t=H$. Otherwise, a
gap truncated at $t=0$ or $t=H$ is counted only fractionally. Since at most two
gaps are affected, $|N(H)-\tilde N(H)|\le 2$, and this error vanishes in the
rate $N(H)/H$ as $H\to\infty$. By ergodicity, the gap rate is therefore
\begin{equation}\label{eq:gaprate}
    \lambda \triangleq \lim_{H\to \infty}\frac{N(H)}{H} =\int_\T\frac{\ind_G(x)}{L(x)}\,d\mu(x).
\end{equation}
Let $N_\ell(H)$ be the number of these gaps with duration at most $\ell$.
Replacing $\ind_G$ by $\ind_G\ind_{\{L\le\ell\}}$ in \eqref{eq:countgap}, the
empirical CDF of the revisit time converges to
\begin{equation}\label{eq:cdf}
    \frac{N_\ell(H)}{N(H)}
    \xrightarrow{H\to \infty}
    F_L(\ell)\triangleq\frac{1}{\lambda}\int_\T\frac{\ind_G(x)\,\ind_{\{L(x)\leq \ell\}}}{L(x)}\,d\mu(x).
\end{equation}
Under this long-term distribution, the mean revisit time is
\begin{equation}\label{eq:meanrevisit}
    \overline{L}\triangleq\lim_{H\to \infty}\frac{1}{N(H)}\int_0^H\ind_G\bigl(\Fl_t(x_0)\bigr)\,dt=\frac{\mu(G)}{\lambda}.
\end{equation}

\textbf{Retransmission Success Probability.} A blindly transmitting IoT device may transmit the same packet multiple times
to increase the probability of successful delivery. If a device transmits at a
random time and is not covered, which it can infer from a missing acknowledgment,
the CDF of the residual waiting time $\tau^+$ satisfies
\begin{multline}\label{eq:residual}
    \p\bigl(\tau^+(\Fl_V(x_0))\leq r \bigm| \Fl_V(x_0)\in G\bigr)\\
    \xrightarrow{H\to \infty}\frac{1}{\overline{L}}\int_0^r\bigl(1-F_L(\ell)\bigr)\,d\ell .
\end{multline}

When designing a retransmission protocol for packets that failed to be
delivered, a relevant quantity is the probability that a retransmission after a
delay $s$ succeeds, given that the first transmission failed,
\begin{equation}\label{eq:probretra}
    p_{\mathrm{re}}(s)\triangleq\lim_{H\to \infty}\p\bigl(\Fl_{V+s}(x_0) \in C\bigm| \Fl_{V}(x_0) \in G\bigr),
\end{equation}
and in particular the delays $s$ that maximize it. For finite $H$,
\begin{multline}\label{eq:retransfinite}
    \p\bigl(\Fl_{V+s}(x_0) \in C\bigm| \Fl_{V}(x_0) \in G\bigr)\\
    =\frac{\frac{1}{H}\int_0^H\ind_G\bigl(\Fl_t(x_0)\bigr)\ind_C\bigl(\Fl_{t+s}(x_0)\bigr)\,dt}{\frac{1}{H}\int_0^H\ind_G\bigl(\Fl_t(x_0)\bigr)\,dt},
\end{multline}
and applying \eqref{eq:spatialaverage} to the numerator and denominator gives
\begin{equation}\label{eq:retrans}
    p_{\mathrm{re}}(s)=\frac{1}{\mu(G)}\int_\T \ind_G(x)\,\ind_C\bigl(\Fl_s(x)\bigr)\,d\mu(x).
\end{equation}

\subsection{Computation}\label{subsec:computation}
The quantities derived in the previous subsection can be evaluated with standard numerical tools. For instance, \eqref{eq:retrans} is a cross-correlation on $\T$ and can therefore be computed for all delays $s$ at once using an FFT, while the gap-related quantities admit a classical flux-based formulation over the boundary $\partial C$. We do not discuss these in detail: any classical two-dimensional numerical integration algorithm suffices. The main advantage over a time-domain simulation lies in convergence and correctness rather than raw computational cost. A simulation samples $\T$ along a single trajectory that advances slowly in $\omega$, completing one cycle per sidereal day, so its statistics converge slowly, in particular for long delays and the tails of the gap distribution. Such simulations cannot simply be extended either, since the accuracy of TLE-based propagation degrades with the time from the element epoch. The integrals over $\T$ instead give the long-run limits directly and independently of any epoch, which also makes parameter sweeps over latitudes and constellation configurations inexpensive.

\section{Numerical Results}\label{sec:numerical}

\begin{table}[!t]
  \centering
  \caption{Parameters of the Numerical Evaluation}
  \label{tab:params}
  \renewcommand{\arraystretch}{1.15}
  \setlength{\tabcolsep}{4pt}
  \begin{tabular}{ll}
    \hline
    \textbf{Parameter} & \textbf{Value}\\
    \hline
    \multicolumn{2}{l}{\textit{Analytical model (Iridium NEXT)}}\\
    Planes $P$ / satellites per plane $S$ & $6$ / $11$ ($T=66$)\\
    Inclination $i$ & $86.7^\circ$\\
    Altitude $h$ & $780$~km\\
    Spacing of adjacent ascending nodes & $31.6^\circ$\\
    Inter-plane phase offset $\Delta$ & $\pi/S$\\
    Integration grid $N_\omega\times N_u$ & $400\times 400$\\
    \hline
    \multicolumn{2}{l}{\textit{SGP4 simulation}}\\
    Satellites (selected / total) & $66$ / $80$\\
    Altitudes & $760$--$800$~km\\
    Sampling interval & $2$~s\\
    Duration & $7$~days\\
    \hline
    \multicolumn{2}{l}{\textit{Terminal and traffic}}\\
    Location (Antwerp, Belgium) & $51.22^\circ$\,N, $4.40^\circ$\,E\\
    Minimum elevation $\varepsilon_{\min}$ & $25^\circ$\\
    First-transmission rate $\nu$ & $1/60~\mathrm{s}^{-1}$\\
    \hline
    \multicolumn{2}{l}{\textit{Design study (Fig.~\ref{fig:constellations})}}\\
    Planes $P$ & $3,\dots,8$\\
    Spacing of adjacent ascending nodes & $\pi/P$\\
    Inter-plane phase offset $\Delta$ & $\pi/S$\\
    Maximum retransmission delay & $3600$~s\\
    \hline
  \end{tabular}
\end{table}

To assess both the correctness of the analysis and the realism of its
assumptions, we evaluate the quantities of Section~\ref{sec:ergodic} for Iridium NEXT, a
Walker Star-like constellation that is a good candidate for Direct-to-Satellite (DtS) communication with IoT devices. We compare the analytical results with
an independent time-domain simulation of the constellation. The Iridium NEXT constellation differs from a
classical Walker Star mainly in two respects: the ascending nodes of adjacent planes are
spaced slightly more than $180^\circ/P$ apart, and not all satellites are at exactly the same altitude. Our analytical model adopts the actual node spacing, i.e., $j\pi/P$ in \eqref{eq:motionsat} is replaced by $j$ times the spacing in Table~\ref{tab:params}, which decreases the maximum revisit time, but it retains a common altitude. All parameters of the model, the simulation and the traffic are listed in Table~\ref{tab:params}. For the design study in Fig.~\ref{fig:constellations}, we keep $i$, $h$ and $\varepsilon_{\min}$ fixed and vary $P$ and $T$.

\subsection{Simulation Methodology}
Of the Iridium NEXT satellites, we select the subset listed in Table~\ref{tab:params}, distributed over six orbital planes of eleven satellites each; the remaining satellites are excluded due to uncertainties regarding their operational status. The selected satellites are propagated with the SGP4 framework~\cite{vallado2006sgp4},
as implemented in the MATLAB Aerospace Toolbox, starting
from their individual TLE epochs. The terminal is placed on the WGS84 ellipsoid in Antwerp, and the elevation is measured
with respect to the ellipsoid normal.

\subsection{Traffic Model}
\label{subsec:traffic}
To validate the protocol-level interpretation of Section~\ref{sec:ergodic}, the
terminal transmits each packet for the first time according to a homogeneous Poisson process with
rate $\nu$, i.e., with independent, exponentially distributed inter-arrival
times of mean $1/\nu$. The Poisson process is only a choice for verification: the same limits hold for any transmission schedule whose timing is independent of the coverage state and not in resonance with the two periods of the constellation, i.e., the in-plane revisit period $2\pi/(Sn)$ and the sidereal day $2\pi/\omega_E$. The results below, therefore, do not depend on the specific traffic model and do not model a complete event-driven packet transmission. A transmission succeeds if the terminal is covered at that instant, i.e., if at
least one satellite is above the mask. If a transmission fails, the terminal retransmits the
packet after a fixed delay $s$, i.e., at $V_i + m s$, $m = 1,2,\dots$, until
the first success. The fraction of successful first transmissions thus estimates the coverage probability~\eqref{eq:covprob}, and the fraction of failed first transmissions whose first retransmission succeeds estimates the conditional probability~\eqref{eq:probretra}, which we compare with its ergodic limit~\eqref{eq:retrans}.

\subsection{Results}
\begin{figure}[!t]
  \centering
  \includegraphics[width=0.95\columnwidth]{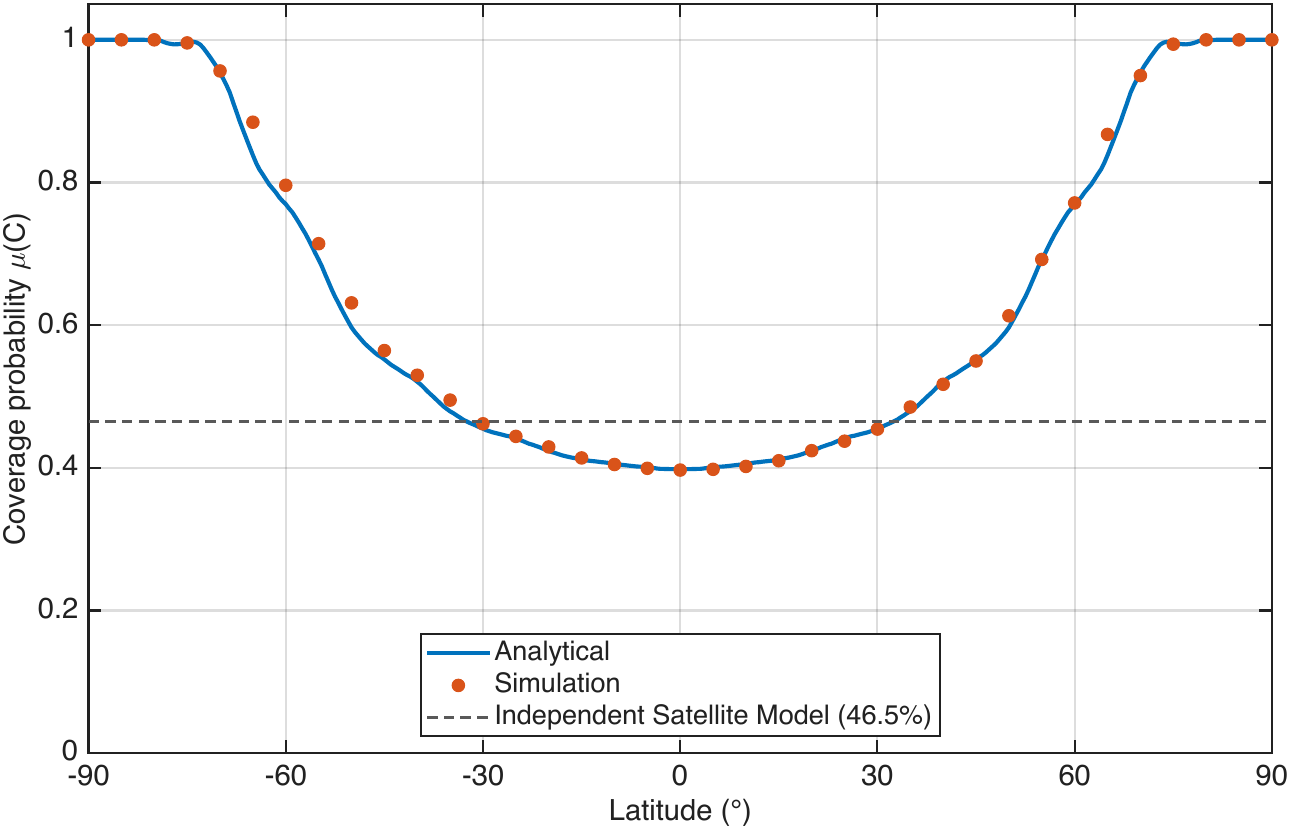}
  \caption{Coverage probability versus terminal latitude: analytical
  $\mu(C)$, MATLAB simulation, and the independent satellite model
  of~\cite{fraire2026blind}.
}
  \label{fig:fractionlat}
\end{figure}

Fig.~\ref{fig:fractionlat} shows the coverage probability as a function of latitude. The analytical curve and the MATLAB simulation agree closely over the full range from $-90^\circ$ to $90^\circ$. We also compare the analytical and simulated values with the independent satellite model of \cite{fraire2026blind}, which approximates coverage as if each satellite were placed independently at random around the Earth. Close to the poles, coverage is continuous: Iridium NEXT is a near-polar Walker Star constellation, and the high density of ascending and descending passes near the poles means that a ground terminal there (almost) never loses coverage. At the equator, the opposite holds: the coverage probability reaches its minimum of about $0.41$, because the orbital planes are farthest apart in this region.
The model in \cite{fraire2026blind} produces a latitude-independent estimate, as any homogeneous binomial or Poisson point process model would. It overestimates the equatorial coverage by about $5$ percentage points, underestimates the coverage in Antwerp by about $15$ percentage points ($0.465$ versus $0.617$), and completely fails to capture the continuous coverage close to the poles. Latitude dependence is therefore an essential feature of a model: a model that ignores it gives inaccurate results for terminals in many regions.

\begin{figure}[!t]
  \centering
  \includegraphics[width=0.95\columnwidth]{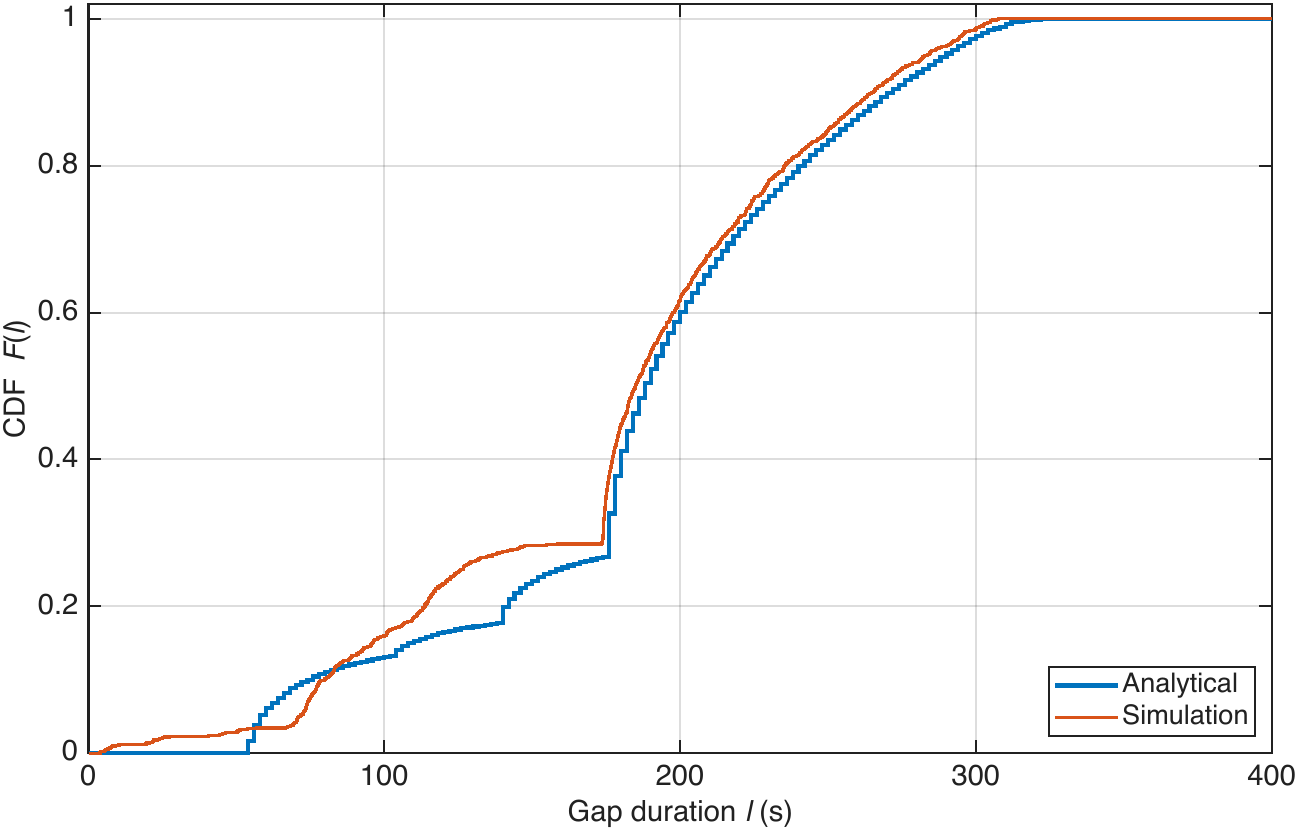}
  \caption{CDF of the revisit time for the terminal in Antwerp: analytical
  $F_L$ from~\eqref{eq:cdf} and empirical CDF of the 1229 gaps in the SGP4 simulation.}
  \label{fig:cdf}
\end{figure}

Fig.~\ref{fig:cdf} shows the CDF of the revisit time for the terminal in Antwerp. The empirical CDF closely matches the analytical one; small deviations appear only for gaps shorter than $200$\,s. Such deviations are expected, since the simulation includes effects that the model neglects, such as the zonal harmonics $J_2$--$J_4$, the ellipsoidal Earth and the small altitude differences between the satellites. The theoretical maximum revisit time of $324$\,s is a relatively accurate estimate of the $308$\,s observed in the simulation. Two effects explain the difference. First, the ellipsoidal Earth shortens the longest gap. Second, the longest gap observed in a finite simulation is only a lower bound on the maximum revisit time, and a longer simulation might be needed to reach it. As noted in Section~\ref{subsec:computation}, however, SGP4 is not suited for much longer simulations, since the accuracy of TLE-based propagation degrades with the time from the element epoch.

\begin{figure}[!t]
  \centering
  \includegraphics[width=0.95\columnwidth]{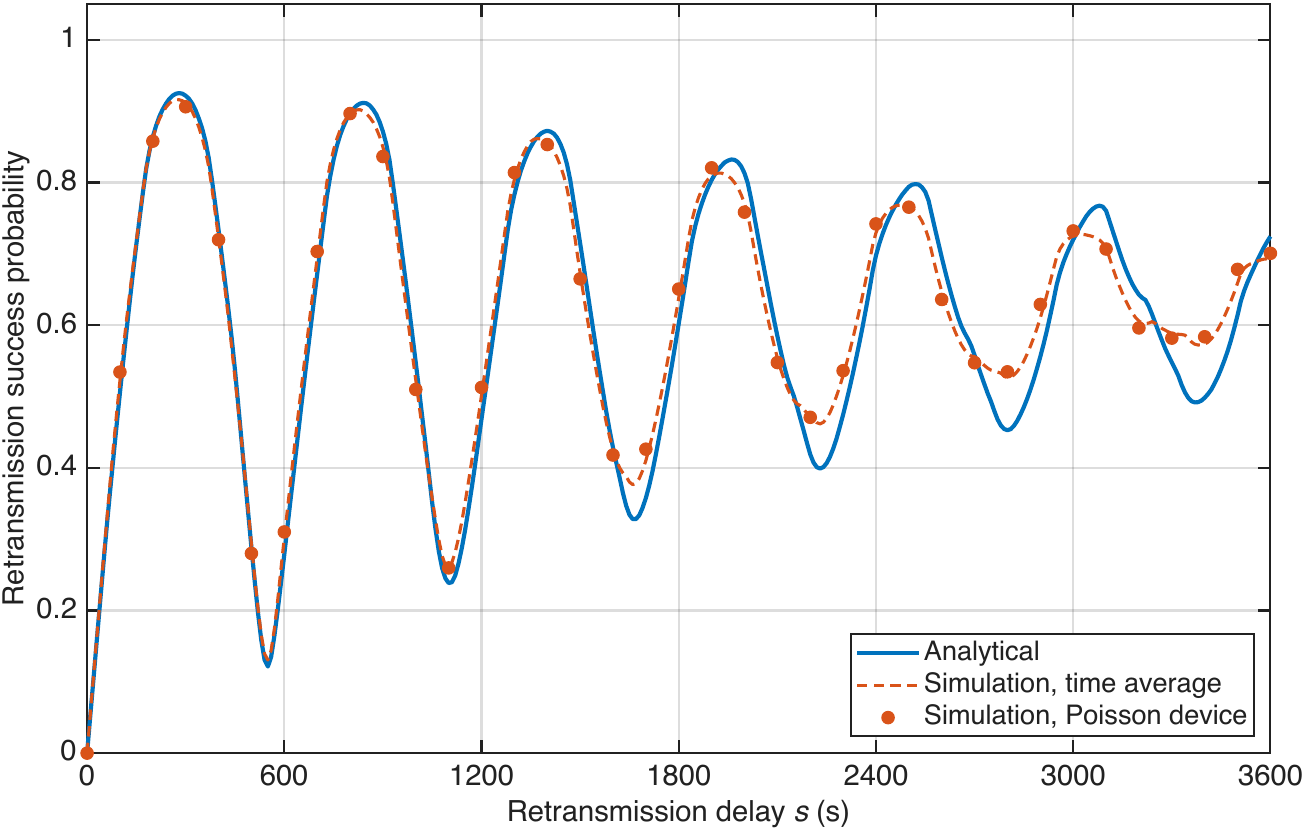}
  \caption{Retransmission success probability $p_{\mathrm{re}}(s)$ versus the
  retransmission delay $s$ for the terminal in Antwerp: analytical
  limit~\eqref{eq:retrans}, time average~\eqref{eq:retransfinite} over the
  simulated coverage intervals, and Poisson-device estimate.}
  \label{fig:retrans}
\end{figure}

Fig.~\ref{fig:retrans} compares the three estimates of $p_{\mathrm{re}}(s)$ for the terminal in Antwerp: the analytical expression~\eqref{eq:retrans}, the finite-window time average~\eqref{eq:retransfinite}, and the fraction of successful first retransmissions in our Poisson traffic model. The curves match closely at short delays and slowly diverge after about an hour. Most importantly, they agree closely on the optimal retransmission delay. At $s=0$, $p_{\mathrm{re}}(s)=0$: a device whose first transmission failed is by definition in a gap, so an immediate retransmission is guaranteed to fail. At $s=280$\,s, the retransmission success probability reaches its maximum of $0.93$. This delay is long enough for the gap in which the first transmission failed to have ended in most cases, but short enough that the next gap has usually not yet started. Knowledge of the temporal structure thus enables good design choices. The curve oscillates with a period close to $2\pi/(Sn)\approx 550$\,s, the time after which each satellite reaches the argument of latitude of the satellite ahead of it in the same plane. After this delay, the state on $\T$ returns to the same $u$ and is shifted by only $2\pi\omega_E/(Sn)\approx 2.3^\circ$ in $\omega$, so a terminal that was in a gap is likely to be in a gap again, which explains the low success probability around this delay. The differences between successive peaks therefore stem from the shift in $\omega$, which accumulates with every period and moves the state across the gap structure of Fig.~\ref{fig:coverage_gap_antwerp}. In summary, Fig.~\ref{fig:retrans} provides direct design guidance: for a terminal in Antwerp, a retransmission timer set to this optimal delay gives the highest success probability among the delays considered, whereas delays in the $400$--$600$\,s range should be avoided. This is particularly important for low-power devices that cannot stay awake continuously.

\begin{figure}[!t]
  \centering
  \includegraphics[width=0.95\columnwidth]{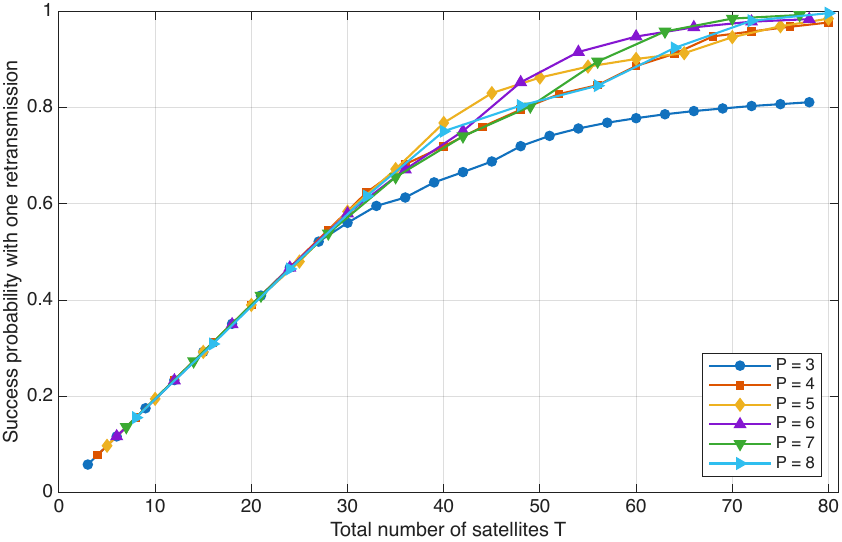}
\caption{Probability $p_{\mathrm{success},1}(s^*)$ that a packet is delivered with at most one retransmission versus the total number of satellites $T$, for Walker Star constellations with different numbers of planes $P$ and the terminal in Antwerp (cf. Table~\ref{tab:params}). For each constellation, the retransmission delay $s^\ast$ maximizes $p_{\mathrm{re}}(s)$ up to the maximum retransmission delay.}
\label{fig:constellations}
\end{figure}

Fig.~\ref{fig:constellations} shows how the framework can be used as a design tool. For Walker Star constellations with different numbers of planes, it shows the probability that a packet is delivered with at most one retransmission, $p_{\mathrm{success},1}(s^*)=\mu(C) + (1-\mu(C))\,p_{\mathrm{re}}(s^*)$, where the retransmission has to be within an hour. IoT devices often lack the storage capacity to store large amounts of data for retransmission many hours later. For the evaluated configurations with up to 30 satellites, the delivery probabilities are nearly identical. For larger values of $T$, constellations with nearly the same total number of satellites can still perform very differently, depending on the number of orbital planes.
\section{Conclusion}\label{sec:conclusion}
We developed an ergodic framework for Walker Star LEO constellations that expresses coverage and temporal statistics as spatial integrals over a two-dimensional torus. The framework captures constellation geometry and provides gap-duration, residual waiting-time, and retransmission statistics relevant to IoT protocol design. The torus construction also applies to other single-shell constellations, including Walker Delta, whose satellites share an altitude and inclination and maintain fixed relative positions.

The results agree closely with a seven-day SGP4 simulation of Iridium NEXT. For the terminal in Antwerp, the optimal retransmission delay among those considered is approximately 280 s, yielding a conditional success probability of 0.93. The constellation comparison also shows that configurations with similar satellite counts can have substantially different delivery probabilities depending on the number of orbital planes.

Future work could incorporate a CINR-based coverage criterion and extend the framework to multi-shell constellations, whose higher-dimensional tori will require more efficient numerical integration. Application-specific studies may also use the framework to guide constellation selection and protocol design.

\bibliographystyle{IEEEtran}
\bibliography{bibliography}

\end{document}